\documentclass[11pt,a4paper]{article}
\usepackage{epsfig}

\begin{document}

\markboth{Yves-Henri Sanejouand}
{A varying-speed-of-light hypothesis can explain supernovae data}

%
%

\title{A simple varying-speed-of-light hypothesis is enough for explaining 
high-redshift supernovae data}

\author{Yves-Henri Sanejouand}
\date{}

\maketitle

\begin{center}
Laboratoire de Physique, UMR 5672 du CNRS,\\
Ecole Normale Sup\'erieure de Lyon,\\
46 all\'ees d'Italie, 69364 Lyon Cedex 07, France.\\
Yves-Henri.Sanejouand@ens-lyon.fr
\end{center}


\begin{abstract}
The hypothesis that the speed of light decreases by nearly
2 cm s$^{-1}$ per year
is discussed within the frame of a simple phenomenological model.  
It is shown that this hypothesis can provide an alternative explanation for the
redshift-distance relationship of type Ia supernovae, which is nowadays 
given in terms of a new form of (dark) energy of unknown origin. 
\end{abstract}

\noindent
Keywords:
cosmological constant;
fine-structure constant;
Hubble law;
lunar ranging laser data;
vacuum permitivity and permeability.

\section{Introduction}

Supernovae can be used as standard candles for cosmological measurements, 
especially those belonging to the rather homogeneous type Ia 
subclass (Sne Ia). In 1998, studies of high-redshift Sne Ia provided 
strong evidences for an acceleration of the universe's 
expansion\cite{Riess:98,Perlmutter:99}, instead of 
the expected deceleration (due to gravitational forces). 
A non-zero cosmological constant
(a "dark energy" with negative pressure)
can explain such results\cite{Riess:98,Perlmutter:99},
but because this explanation bears greatly on "new physics"\cite{Peebles:03}
and involves a "cosmic coincidence"\cite{Zlatev:99,Sahni:02},
alternative ones have to be explored.
For instance, from an epistemological point of view, a non-zero cosmological constant
would mean that laws of physics at cosmological scales are different from
what can be observed on earth, breaking down one of the major 
principle followed by physicists with so many successes over 
centuries. 

\noindent
The purpose of this paper is to show that high-redshift supernovae
data are consistent with the hypothesis that the speed of light is
time-dependent. Such an hypothesis was already proposed during the 
thirties, for explaning the cosmological redshift\cite{Takehuchi:31,Wold:35,North}.
It has been reconsidered more recently\cite{Tan:86},
noteworthy within the frame of standard cosmological models,
although not on the same timescale\cite{Magueijo:00,Ranada:03};
for a discussion of the status of present varying-speed-of-light theories, 
see Ref. \cite{Magueijo:03}.

\begin{figure}[t]
\begin{center}
 \label{sn}
 \epsfig{width=8cm,angle=270,figure=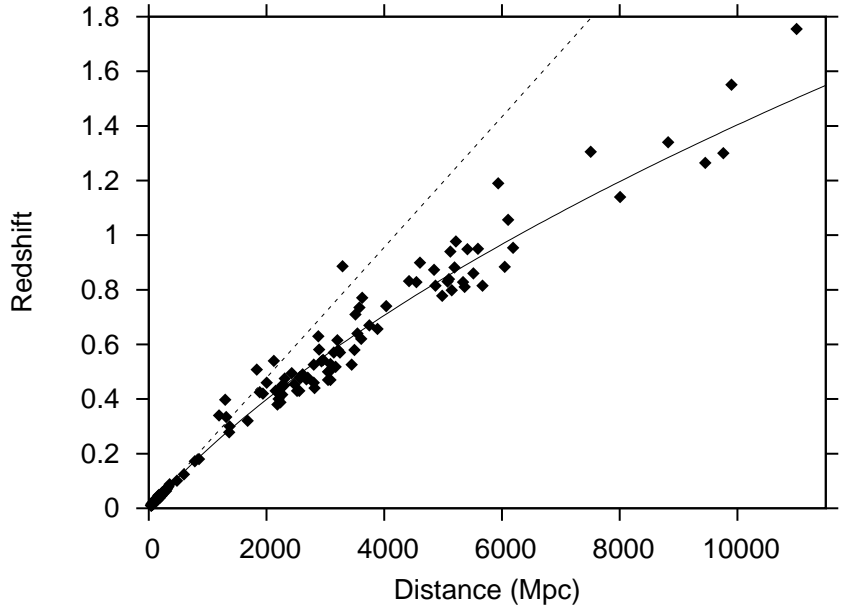}
 \caption{Redshift of type Ia supernovae as a function of their distance.
 Black diamonds: the 156 supernovae of the recently compiled "gold set".
 Dotted line: Hubble law. Plain line: the relationship expected
 according to the varying-speed-of-light hypothesis
 discussed in the present study. This is {\it not} a fit of the 
 data. The distance scale is set using a value
 for the Hubble constant of
 72 km s$^{-1}$ Mpc$^{-1}$.}
\end{center}
\end{figure}

\section{Redshift-distance relationship}

Light curves of distant Sne Ia are dilatated in time\cite{Perlmutter:96,Riess:96},
according to: 
\begin{equation}
\label{aging}
\frac {T_d}{T_0} = 1 + z 
\end{equation}
where $T_0$ and $T_d$ are the typical timescales of the event, as observed in
the case of nearby and distant Sne Ia, respectively, $z$ being the redshift
of the distant supernova.
As a matter of fact, a stretching by a $(1 + z)$ 
factor of reference, nearby, Sne Ia light curves is included
in all analyses of distant Sne Ia data\cite{Perlmutter:96,Riess:96}.

\noindent
Within the frame of standard cosmological models, (\ref{aging}) is understood
through the hypothesis that the corresponding length ($c_0 T_0$)
increases as a function of time,
$c_0$, the speed of light, 
being assumed to have the same, time-independant, value, as measured on earth\cite{Schrodinger:39,Wilson:39}.
Let us assume instead that (\ref{aging}) reflects the fact that the speed
of light is time-dependent, so that:
\begin{equation}
\label{zc}
\frac {c_d}{c_0} = 1 + z 
\end{equation}
where $c_d$ is the speed of light at time
$t_d$, when the photons were emitted.
Indeed, a relationship between a timescale measured by an observer
and the speed of light at the time it was emitted, so that
$T_d \propto c_d$
can be obtained in the context of non-relativistic models
(see Ref. \cite{Wold:35}, as an early example).
Hereafter, for the sake of clarity,
the chosen point of view is also non-relativistic. However,
note that an effective variation of the speed of light, as a 
consequence of a variation of the gravitational field,
can also be obtained within the frame of General Relativity\cite{Takehuchi:31,Magueijo:03}.

\noindent
If the speed of light varies slowly as a function 
of time, $c_d$ can be approximated by:
\[                    
c_d = c_0 + a_c \Delta t + \frac{1}{2} \dot{a}_c \Delta t^2 +
\cdots 
\]
where $a_c$ is the time derivative of the speed of light, 
$\dot{a}_c$ the time derivative of $a_c$, 
and where $\Delta t$ is the photon time-of-flight between its source and the observer.
Let us consider
only the first terms of this expansion, namely, that for small
enough values of $\Delta t$:
\begin{equation}
\label{ct}
c_d = c_0 + a_c \Delta t 
\end{equation}
During $\Delta t$, the photon travels along a path of length $d$, so
that:
\[
d = c_0 \Delta t + \frac {1}{2} a_c \Delta t^2 
\]
This yields:
\begin{equation}
\label{dt}
\Delta t = \frac {c_0} {a_c}
(\sqrt{ 1 + \frac {2 a_c d}{c_0^2}} - 1)
\end{equation}
Thus, with (\ref{ct}):
\begin{equation}
\label{cd}
c_d = c_0 \sqrt{ 1 + \frac {2 a_c d}{c_0^2}}
\end{equation}
and, with (\ref{cd}) in (\ref{zc}):
\begin{equation}
\label{zd}
z = \sqrt{ 1 + \frac {2 a_c d}{c_0^2}} - 1
\end{equation}
For short distances, 
(\ref{zd}) can be approximated by:
\begin{equation}
\label{newhub}
z = \frac {a_c d}{c_0^2}  
\end{equation}
a relationship of the same form as Hubble law, namely:
\begin{equation}
\label{hub}
z = \frac {H_0 d}{c_0}  
\end{equation}
where $H_0$ is the Hubble constant.
So, with (\ref{newhub}) and (\ref{hub}):
\begin{equation}
\label{ac}
a_c = H_0 c_0  
\end{equation}
and (\ref{zd}) becomes:
\begin{equation}
\label{zdnop}
z = \sqrt{ 1 + \frac {2 H_0 d}{c_0}} - 1
\end{equation}
Note that (\ref{zdnop}) can also be viewed as a straightforward generalisation
of Hubble law. Indeed, rewriting (\ref{hub}) as
$z = H_0 \Delta t$,
(\ref{zdnop}) is then obtained from (\ref{dt}) and (\ref{ac}). 
In other words, (\ref{zdnop}) is a modified version of Hubble law
where $\Delta t$, the photon time-of-flight, is calculated so as to
take into account the time-dependence of the speed of light.

\noindent
Interestingly, (\ref{ac}) has already been obtained in other contexts.
First, as the critical value for the acceleration below which Newton laws
are no longer valid, according to the MOND alternative explanation of 
the dark matter problem\cite{Milgrom:83}. Then, 
as the value of an 
anomalous, unexplained, acceleration directed towards the sun,
that has been found to act on distant spacecrafts in the solar
system, noteworthy Pioneer 10 and 11, launched some thirty years 
ago\cite{Anderson:98,Anderson:02}.

\section{Data}
The 156 supernovae considered in this study are those taken into account
within the frame of an analysis of 16 high-redshift supernovae observed
with the Hubble Space Telescope; see Table 5 in Ref. \cite{Riess:04}.
This "gold set" has the virtue that all distance estimates
were derived from a single set of algorithms\cite{Riess:04}.
In order to check (\ref{zdnop}) against these experimental data, 
magnitudes were first translated into actual distances, 
as it is legitimate to do in the case of standard candles.
However, assuming a given value for $H_0$ is still necessary
due of the lack of enough reliable data in the case of nearby Sne Ia.

\section{Results}
In Figure \ref{sn}, Sne Ia redshifts are plotted as a function
of their distance.
The plain line is {\it not} a least-square fit
of these data. It corresponds to (\ref{zdnop}),
that is,
to the hypothesis that Sne Ia redshifts 
are observed
as a consequence of the time-dependence of the speed of light.
More specifically, according to (\ref{ac}), if 
$H_0 = 72$ km s$^{-1}$ Mpc$^{-1}$
\cite{Freedman:01}, then
$a_c = 7~10^{-10}$ m s$^{-2}$. 
This corresponds
to a change in the speed of light of 
-2.2 cm s$^{-1}$ per year.

\section{Discussion}

\subsection{Fine-structure constant}

Given the central role of the speed of light in modern theoretical physics,
consequences of its variation, even at slow rate, are far reaching.
But from an experimental point of view the major constraint comes from the
fact that 
$\alpha = \frac{e^2}{4 \pi \epsilon_0 \hbar c_0}$,
the fine-structure constant,
depends very little upon the redshift\cite{Webb:99},
if it does at all\cite{Uzan:03,Petitjean:04}. This means
that if the speed of light varies in time as much as assumed herein, 
then either $\epsilon_0$, the vacuum permitivity, $e$,
the electron charge, $h$, the Planck constant, or both, vary
as well. However, because there is a known link between $c_0$ and $\epsilon_0$,
namely: 
\begin{equation}
c_0=\frac{1}{\sqrt{\mu_0 \epsilon_0}} 
\label{cvac}
\end{equation}
where $\mu_0$
is the vacuum permeability, the simplest hypothesis is to assume that the vacuum permitivity also
varies in time, so that:
\begin{equation}
\epsilon(t) c(t) = \frac{e^2}{4 \pi \hbar \alpha}
\label{acst}
\end{equation}
does not. Moreover, (\ref{cvac}) and (\ref{acst}) are consistent only if the vacuum permeability
also varies in time, so that:
\[
\mu(t) = \frac{(4 \pi \hbar \alpha)^2}{e^4} \epsilon(t)
\] 
In other words, 
if the speed of light varies in time,
while the fine-structure constant does not,
then the ratio between 
vacuum permitivity and permeability also does not.
However, building a self-consistent theory of the relationship between light, matter
and vacuum properties, as done for instance in another context in Ref. \cite{Ranada:03},
is beyond the scope
of this study. As a matter of fact, such a kind of work may await confirmation
at the experimental level, as well as further clues, in order to be developped on firm 
enough grounds. 

\subsection{Physical units}

In the present international unit system, the value
of $c_0$ is exactly 299,792,458 m s$^{-1}$, {\it par d\'efinition}. 
But because $c_0$ is involved in
other physical units, 
variations of some of the corresponding physical constants should
reflect any actual time-dependence of the speed of light. However, most physical constants
are known with a relative standard uncertainty of $10^{-9}$,
according to the 2002 CODATA set of recommended values. This is
likely to be not yet enough for demonstrating a relative variation
of $10^{-10}$ per year, as expected herein for the speed of light.

\subsection{Lunar laser ranging}

Distances in the Solar system can be measured using radar or laser impulses,
the Earth-Moon distance being the most accurately determined one\cite{Dickey:94}.
So, if the speed of light varies in time, a systematic trend should
be observed in these distance measurements.
Indeed, such a trend has been obtained 
by laser ranging. It corresponds to a rate of 
change of the measured Earth-Moon distance of
3.82 $\pm$ 0.07 cm per year\cite{Dickey:94}.
These distance time series, $d_{mes}(t)$, are obtained assuming that:
\begin{equation}
\label{dmes}
d_{mes}(t) = \frac {c_0 \delta t}{2}
\end{equation}
where $\delta t$ is the time taken by light to go to the Moon and back to
to the observer. If the speed of light varies in time,
$\delta t$ is approximatively given by: $\delta t = \frac{2 d_0} {c(t)}$,
where $d_0$ is the actual Earth-Moon distance, that is, with (\ref{dmes}):
\[
d_{mes}(t) = \frac{c_0}{c(t)} d_0
\]
According to the present study, $c(t)$ is given by (\ref{ct}), with $t=0$
when the first distance measurements were performed. Taking into account (\ref{ac}) yields:
\[
d_{mes}(t) = \frac{d_0}{1 - H_0 t}
\]
So, as a consequence of the variation of the speed of light, 
the rate of change of the measured Earth-Moon distance, 
$v_{mes}$, 
is expected to be,
for short periods of measurements (accurate measurements of the Earth-Moon distance
have been performed over the last thirty years, after reflectors were let on the
Moon by Apollo missions): 
\[
v_{mes} = H_0 d_0
\]
that is, a Hubble-like relationship. In other words, according to the present study,
distances in the solar system should seem to increase in time, as if the solar system
were expanding at the rate of universe's expansion. 
So, depending upon the actual value of $H_0$, part or all 
the lunar recession measured by
laser ranging could be due to the time-dependence of the speed of light.
In particular, if $H_0 = 97$ km s$^{-1}$ Mpc$^{-1}$ it is an apparent effect. 

\noindent
On the other hand if 
$H_0 = 72$ km s$^{-1}$ Mpc$^{-1}$\cite{Freedman:01},
the pseudo-lunar recession due to the variation of the speed of light is of 2.8 cm per year.
The nearly 1 cm per year difference could come from an actual lunar 
recession, as a consequence of tidal forces.
Such forces are also expected to be responsible for a
secular change in the length of the day (LOD). 
In a remarkable compilation of anciant eclipses\cite{Stephenson:95},
it was shown that the mean LOD change has been of
+1.70 $\pm$ 0.05 milliseconds per century over the last 2500 years. 
Under the hypotheses that the momentum of the Moon-Earth system
is conserved and that, during this period, all the LOD change was due to tidal forces,
the tidally-driven lunar recession should have been 
of nearly 2.8 cm per year\cite{Stephenson:95}, leaving a value 
for the pseudo-lunar recession 
of only 1 cm per year.
However, significant LOD fluctuations
are observed in the anciant eclipses data, on the millenium timescale\cite{Stephenson:95},
while fluctuations of several milliseconds have been observed over the last centuries,
likely to be due to events like
the warm El Nino Southern Oscillation, which is accompanied
by an excess in atmospheric 
angular momentum\cite{Munk:02}.
As a matter of fact, between 1969 and 2005,
that is, 
while the lunar laser ranging data were collected,
the mean LOD has decreased.

\section{Conclusion}

From a theoretical point of view.
the varying-speed-of-light hypothesis discussed in this study is challenging.
However, because it is consistent with experimental data,
and especially because it explains the supernovae data so well, 
it should prove useful, at least at an heuristic level.
From an experimental point of view, measuring the speed of light
with a cm s$^{-1}$ accuracy several years in a row may not be
out of reach. However, measuring its variations at this
level of accuracy should prove easier. 


\label{lastpage}


\end{document}